\documentclass[12pt]{article}
\usepackage[utf8]{inputenc}
\usepackage{amsmath, amsthm, amssymb, amsfonts}
\usepackage[numbers, comma]{natbib}

\usepackage{fullpage}
\usepackage{authblk}
\usepackage[bibliography=common]{apxproof}
\usepackage{import}
\usepackage{mathtools}
\usepackage{tikz}
\usepackage{float}
\usepackage{enumitem}
\usetikzlibrary{calc,patterns,angles,shapes, positioning, intersections, quotes}
\newtheorem{theorem}{Theorem}
\newtheoremrep{thm}{Theorem}
\newtheoremrep{lem}{Lemma}

\usepackage{booktabs}      
\usepackage{multirow} 
\usepackage{makecell}      

\usepackage{graphicx}
\newtheorem{corollary}[theorem]{Corollary}

\theoremstyle{definition}
\newtheorem{definition}{Definition}

\theoremstyle{example}

\usepackage[colorlinks=false,pagebackref=true, hidelinks]{hyperref}

\newcommand{\R}{\mathbb{R}}

\def\C{\mathcal{C}}
\def\V{\mathcal{V}}
\def\F{\mathcal{F}}

\def\th{\theta}

\def\E{\mathbb{E}}

\allowdisplaybreaks

\date{\today\\\small{\href{https://goelsumit.com/files/contests_finitetype.pdf}{(Link to latest version)}}}

\begin{document}

\title{The effect of competition in contests: A unifying approach
\thanks{We are grateful to Olivier Bochet, Federico Echenique, Wade Hann-Caruthers, Christian Seel, Forte Shinko, as well as conference participants at Ashoka University, Delhi School of Economics, NYU Abu Dhabi, Stony Brook University, and University of Reading for helpful comments and suggestions. Financial support by the Center for Behavioral Institutional Design and Tamkeen under the NYU Abu Dhabi Research Institute Award CG005 is gratefully recognized. An earlier version of this paper circulated under the title ``Contest design with a finite type-space: A unifying approach."}}

\author{Andrzej Baranski\thanks{NYU Abu Dhabi; \href{mailto:a.baranski@nyu.edu}{a.baranski@nyu.edu}} \quad Sumit Goel\thanks{NYU Abu Dhabi; \href{mailto:sumitgoel58@gmail.com}{sumitgoel58@gmail.com}; 0000-0003-3266-9035}}

\maketitle

\begin{abstract}
We study how increasing competition, by making prizes more unequal, affects effort in contests. In a finite type-space environment, we characterize the equilibrium, analyze the effect of competition under linear costs, and identify conditions under which these effects persist under general costs. Our findings reveal that competition may encourage or deter effort, depending on the relative likelihood of efficient versus inefficient types. We derive implications for the classical budget allocation problem and establish that the most competitive winner-takes-all contest is robustly optimal under linear and concave costs, thereby resolving an open question. Methodologically, our analysis of the finite type-space domain---which includes complete information as a special case and can approximate any continuum type-space---provides a unifying approach that sheds light on the contrasting results in these extensively studied environments.


\end{abstract} 
\section{Introduction}

Contests, which reward agents with prizes based on their relative performance, are widely used to encourage effort in various settings, such as crowdsourcing for innovation, sporting events, and classrooms. Given their prevalence, understanding how different contest structures influence agents’ effort is crucial. A natural question in this context is how increasing the competitiveness of a contest, by making prizes more unequal, affects agents' incentives to exert effort. In recent work, \citet*{fang2020turning} examined this question in a complete information environment with homogeneous agents, uncovering a clear monotonic relationship driven by the shape of the effort cost function: increasing competitiveness encourages effort under concave costs and discourages effort under convex costs. However, the assumption of complete information is quite restrictive, as in many real-world contexts, agents posses privately-known abilities. \\

In this paper, we analyze the effect of increasing competitiveness on effort in an environment where agents have private abilities drawn independently from a finite type-space. Given a contest and their private types, agents simultaneously choose their effort levels, incur associated (type-dependent) costs, and receive prizes based on the rank order of their efforts. We first characterize the unique symmetric equilibrium of the Bayesian game. The equilibrium involves different agent types mixing over disjoint but continuous intervals, such that more efficient types outperform less efficient types with probability one. We then analyze the effect of increasing competition on expected equilibrium effort under linear effort costs, and identify conditions under which these effects persist under general costs. \\

Our findings establish that under linear and concave costs, increasing competitiveness by transferring value to the best-ranked prize always encourages effort. Consequently, for the classical contest design problem of allocating a fixed budget across prizes, the winner-takes-all contest is uniquely optimal under linear and concave costs. While \citet*{moldovanu2001optimal} previously derived a similar result in a continuum type-space environment, the optimality of awarding multiple prizes in complete information environments (\citet*{barut1998symmetric, clark1998competition, cohen2008allocation}) led \citet*{sisak2009multiple} to conjecture that in our (intermediate) finite type-space setting, multiple prizes might still be optimal:
\begin{quote}\textit{``The case of asymmetric individuals, where types are private information but drawn from discrete, identical or maybe even different distributions, has not been addressed so far. From the results ... on asymmetric types with full information, one could conjecture that multiple prizes might be optimal even with linear costs."}
\end{quote}

Our analysis demonstrates that, at least when types are drawn from identical distributions, this conjecture does not hold. As soon as there is any little uncertainty in the environment, the winner-takes-all contest is robustly optimal under linear and concave costs. \\

In general, however, increasing competitiveness by transferring value to better-ranked prizes (other than the best-ranked prize) may encourage or discourage effort, depending upon the underlying distribution of types. Intuitively, given the monotonic structure of the equilibrium, increasing competition reduces the expected equilibrium prize of the less efficient types while raising it for the more efficient types. Unlike the complete information case---where this effect is fully captured by the changes in effort costs, as equilibrium utilities are zero---the transformation additionally impacts the equilibrium utilities (or information rents, in the language of mechanism design) of the more efficient types in a setting with private types. We explicitly compute these effects on equilibrium utilities and expected effort under linear costs and show that as long as the equilibrium utility of the most efficient type does not increase, the effects on effort under linear costs are informative about those on equilibrium effort under general costs: the non-negative effects extend to concave costs, while the non-positive effects extend to convex costs. This result generalizes the complete information analysis of \citet*{fang2020turning}, where increasing competition has no effect on equilibrium utility or expected effort under linear costs. In a binary type-space environment with linear costs, we show that increasing competition always encourages effort when the efficient type is sufficiently likely, while it always discourages effort (except when transferring value to the best-ranked prize) when the inefficient type is sufficiently likely.\\

The existing literature in contests has predominantly focused on the design problem in environments where the type-space is either a continuum, or a singleton (the complete information case), and the results highlight how the structure of the optimal contest can vary significantly depending on the environment. For the continuum type-space, the most competitive winner-takes-all contest has been shown to be optimal under linear or concave costs (\citet*{moldovanu2001optimal}), in some cases under convex costs (\citet*{zhang2024optimal}), with negative prizes (\citet*{liu2018optimal}), and with general architectures (\citet*{moldovanu2006contest, liu2014effort}). In comparison, in the complete information environments, the minimally competitive budget distribution (all agents but one receive an equal positive prize) has been shown to be a feature of the optimal contest quite generally (\citet*{barut1998symmetric, letina2023optimal, letina2020delegating, xiao2018equilibrium}). In a general framework with many agents, \citet{olszewski2016large, olszewski2020performance} show that awarding multiple prizes of descending sizes is optimal under convex costs. Other related work has examined the effect of competition in complete information setting (\citet*{fang2020turning}), and continuum type-space setting (\citet*{goel2023optimal, krishna2024pareto}). The finite type-space environment embeds the complete information as a special case and can approximate any continuum type-space. Thus, our analysis of this intermediate and fundamental domain not only bridges a gap in the literature, but provides a unifying approach offering insights into the contrasting results in these extreme environments.\footnote{In early work, \citet*{glazer1988optimal} highlight this distinction by solving the problem in some special cases. Other related studies include \citet*{schweinzer2012optimal, drugov2020tournament} who examine the budget allocation problem under different contest success functions. For general surveys of the literature in contest theory, see \citet*{corchon2007theory, sisak2009multiple, dechenaux2015survey, vojnovic2015contest,konrad2009strategy, chowdhury2023heterogeneity, fu2019contests, bevia2024contests}.}\\

There is a related literature on contests with a finite type-space, much of which assumes binary type-spaces or a small number of agents and focuses on characterizing equilibrium properties under asymmetric or correlated types. \citet*{siegel2014asymmetric} establishes the existence of a unique equilibrium under general distributional assumptions. With correlated types, \citet*{liu2016symmetric} show that the symmetric equilibrium may be non-monotonic when the degree of absolute correlation is high, \citet*{rentschler2016two} highlight the possibility of allocative inefficiency in equilibrium, while \citet*{tang2023reservation} and \citet*{kuang2024ridge} explore the impact of reservation prices and information disclosure policies, respectively. With asymmetric type distributions, \citet*{szech2011asymmetric} shows that agents may benefit from revealing partial information about their private types to their opponents. Taking this information to be exogenous, \citet*{chen2021all} characterizes equilibrium outcomes for varying levels of signal informativeness. Other contributions include \citet*{konrad2004altruism}, who examines contests with altruistic or envious types, and \citet*{einy2017common}, who analyze common value all-pay auctions with private asymmetric information. In contrast to this literature, our paper assumes symmetric and independent types and focuses on studying the effect of competition on effort with arbitrarily general finite type-spaces.\footnote{In other related work, \citet*{ewerhart2020unique} study imperfectly discriminating contests with a finite type-space, identifying conditions for uniqueness of equilibrium. There is also some work in mechanism design, and auction design with finite type-spaces (\citet*{maskin1985auction,jeong2023first, vohra2012optimization,lovejoy2006optimal, doni2013revenue, elkind2007designing}).}

\section{Model}

\subsection*{Contest environment}
There is a set of $N+1$ risk-neutral agents. Each agent has a privately known type, which is its effort cost function. We assume that there are $K$ possible types, each of which is such that agents incur zero cost from zero effort, higher cost from higher effort, and arbitrarily large costs from arbitrarily large effort. Formally, each agent's private type is drawn from a finite type-space $$\C=\{c_k \in \F: k \in [K]\},$$ where the set $\F$ is defined as
$$\F=\{c: \R_+\to \R_+ \mid c(0)=0, c'(x)>0 \text{ for all } x>0, \text{ and } \lim_{x\to \infty} c(x)=\infty\}.$$

We further assume that the $K$ possible types can be ordered by efficiency, and without loss of generality, let types associated with higher indices be more efficient than those with lower indices. Formally, the type-space $\C$ is an \textit{ordered type-space}, defined as follows:

\begin{definition}[Ordered type-space]
\label{ass:ordered_types}
A type-space $\C=\{c_k \in \F: k \in [K]\}$ is an \textit{ordered type-space} if, for all $x>0$, $$c_1'(x)>\dots >c_K'(x).$$
\end{definition}

A particularly relevant subclass of ordered type-spaces, commonly studied in the literature on contests (with a continuum of types), consists of type-spaces where the types are simply scaled versions of a single base function.

\begin{definition}[Parametric type-space]
A type-space $\C=\{c_k \in \F: k \in [K]\}$ is a \textit{parametric type-space} if there exists a (base) cost function $c\in \F$ and parameters $\th_1, \dots, \th_K \in \R_+$, with $\th_1>\dots>\th_K$, such that for each $k \in [K]$, $$c_k(x)=\th_k c(x) \text{ for all }x\in \R_+.$$
    
\end{definition}

Each agent's private type is drawn independently from the type-space $\C$ according to distribution $p=(p_1, \dots, p_K)$, where $p_k>0$ for all $k$ and $\sum_{k=1}^K p_k=1$. For each $k$, we let $P_k=\sum_{j=1}^k p_j$. We refer to the collection $(N+1, \C, p)$ as the \textit{contest environment} and assume that it is common knowledge.

\subsection*{Contest}

A \textit{contest} $v=(v_0, \dots, v_N)$ assigns a prize value to each rank, with $v_0 \leq \dots \leq v_N$ and $v_0<v_N$. Given a contest $v$ and their private types, all $N+1$ agents simultaneously choose their effort. The agents are ranked according to their efforts, with ties broken uniformly at random, and awarded the corresponding prizes. Specifically, the agent who exerts the highest effort (outperforming all other $N$ agents) is awarded the prize $v_N$, and more generally, the agent who outperforms exactly $m \in \{0, \dots, N\}$ out of the $N$ other agents is awarded the prize $v_m$. If an agent of type $c_k\in \C$ wins prize $v_m$ after exerting effort $x_k\geq 0$, their payoff is equal to the value of the prize minus the cost of exerting the effort:
$$v_m-c_k(x_k).$$

Given a contest environment $(N+1, \C, p)$, a contest $v$ defines a Bayesian game between the $N+1$ agents. Since the game induced by $v$ is strategically equivalent to the game induced by the contest $w$ where $w_m=v_m-v_0$ for all $m\in \{0, \dots, N\}$, we assume without loss of generality that $v_0=0$. Formally, we will restrict our attention to contests in the set $$\V=\{v\in \R^{N+1}: v_0\leq v_1 \leq \dots \leq v_N \text{ where } 0=v_0<v_N\}.$$

\subsection*{Equilibrium}
For any contest environment $(N+1, \C, p)$ and contest $v\in \V$, we will focus on the symmetric Bayes-Nash equilibrium of the induced Bayesian game. This is a strategy profile where all agents use the same (potentially mixed) strategy, mapping types to a distribution over non-negative effort levels, such that if an agent has type $c_k$, choosing any effort level in the support of the distribution for $c_k$ yields an expected payoff at least as high as any other effort level, given that all other agents use the same strategy. We denote this symmetric Bayes-Nash equilibrium by $(X_1, X_2, \dots, X_K)$, where $X_k\sim F_k$ represents the random level of effort exerted by an agent of type $c_k$. We further denote by $X\sim F$ the ex-ante random level of effort exerted in equilibrium by an arbitrary agent, so that for any $x\in \R$, 
$$
F(x)= \sum_{k=1}^K p_kF_k(x),    
$$
and the expected effort of an arbitrary agent is $$
\E[X]=\sum_{k=1}^K p_k\E[X_k].
$$

\subsection*{Competition}

We are interested in examining how increasing competitiveness of a contest influences the expected equilibrium effort. As is standard in the literature, we define a contest as being more competitive than another if the prizes are more unequal, measured using the Lorenz order.

\begin{definition}
A contest $v\in \V$ is \textit{more competitive} than $w\in \V$ if $v$ is more unequal than $w$ in the Lorenz order, i.e., 
$$\sum_{i=0}^m v_i \leq  \sum_{i=0}^m w_i \text{ for all } m\in \{0, 1, \dots, N\},$$
with equality for $m=N$.
\end{definition}

Observe that, given a fixed budget $V \in \R_+$, the contest that awards the entire budget to only the best-performing agent, $v=(0, 0, \dots, 0, V)$, is more competitive than any other contest $w\in \V$ that distributes the entire budget. At the other extreme, the contest that distributes the budget equally among all but the worst-performing agent,  $v=(0, \frac{V}{N}, \dots, \frac{V}{N})$, is less competitive than any other contest $w\in \V$ that distributes the entire budget. \\

Importantly, if $v \in \V$ is more competitive than $w\in \V$, 
$v$ can be obtained from $w$ through a sequence of transfers from lower-ranked prizes to higher-ranked prizes. The marginal effect of such a transfer, say from prize $m'$ to $m$ with $m>m'$, is captured by
$$\dfrac{\partial \E[X]}{\partial v_m}-\dfrac{\partial \E[X]}{\partial v_{m'}}.$$
Our objective is to evaluate the impact of increasing competition on expected equilibrium effort across different contest environments, and consequently, identify features under which it may encourage effort, as well as those under which it may discourage effort. We will explore implications of these findings for the classical design problem of allocating a fixed budget across different prizes to maximize expected equilibrium effort.

\subsection*{Notation}
We now introduce some notation used throughout the rest of the paper. We let
$$H^N_{m}(t)={N \choose m} t^m(1-t)^{N-m}$$ denote the probability that a binomial random variable $Y\sim Bin(N, t)$ takes the value $m$. 
We also let
$$H^N_{\leq m}(t)=\sum_{i=0}^m H^N_i(t)\text{ and }H^N_{\geq m}(t)=\sum_{i=m}^N H^N_i(t),$$ denote the probabilities that $Y\sim Bin(N, t)$ takes a value at most $m$ and at least $m$, respectively.\\

Given a contest $v\in \V$, note that if an agent outperforms each of the $N$ other agents independently with probability $t\in [0,1]$, then $H^N_m(t)$ represents the probability that the agent outperforms exactly $m$ out of these $N$ agents, thereby being awarded the prize $v_m$. We thus define
$$\pi_v(t)=\sum_{m=0}^{N} v_m H^{N}_{m}(t),$$ and note that it denotes the expected value of the prize that an agent is awarded if it outperforms each of the $N$ other agents independently with probability $t\in [0,1]$.

\section{Equilibrium}

In this section, we characterize the symmetric Bayes-Nash equilibrium of the Bayesian game. Before providing a complete description, we establish a robust structural property of this equilibrium: agents mix over contiguous intervals, with more efficient agents choosing greater effort than less efficient agents.

\begin{lemrep}
\label{eqbm_intervals}
Consider any contest environment $(N+1, \C, p)$ where $\C$ is an ordered type-space. For any contest $v\in \V$, a symmetric Bayes-Nash equilibrium $(X_1, \dots, X_K)$ must be such that there exist boundary points $b_0<b_1< \dots< b_K$, with $b_0=0$, so that for each $k \in [K]$, $X_k$ is continuously distributed on $[b_{k-1}, b_k]$.
\end{lemrep}
\begin{proof}
Suppose $(X_1, X_2, \dots, X_K)$ is a symmetric Bayes-Nash equilibrium, and let $X\sim F$ denote the ex-ante effort of an arbitrary agent. Let $u_k$ denote the payoff of agent of type $c_k\in \C$ under this symmetric strategy profile.

\begin{enumerate}
    \item \textbf{Mixed strategies: }We first show that $X_k$ cannot have any atoms. Suppose instead that $\Pr[X_k=x_k]>0$. We will argue that an agent of type $c_k$ obtains a strictly higher payoff from choosing $x_k+\epsilon$ as compared to $x_k$ for $\epsilon>0$ and small enough. Notice that under the given profile, there is a positive probability that all $N+1$ agents are tied at effort level $x_k$, in which case the ties are broken uniformly at random.  Thus, choosing $x_k+\epsilon$ results in a discontinuous jump in the expected value of the prize awarded to the agent (since $v_0<v_N$), even though the additional cost $c_k(x_k+\epsilon)-c_k(x_k)$ can be made arbitrarily small with $\epsilon$ small enough. It follows that for agent of type $c_k\in \C$, the payoff from choosing $x_k+\epsilon$ is strictly higher than that from choosing $x_k$, which is a contradiction. Thus, $X_k$ must be a continuous random variable. Consequently, we assume, without loss of generality, that the support of $X_k$ is closed.

    \item \textbf{Disjoint support (essentially) across types:} We now show that for any $j\neq k$, the support of $X_j$ and $X_k$ have at most one point of intersection. Suppose instead that both $x,y$ are in the support of both $X_j$ and $X_k$ and $x\neq y$. Since an agent of type $c_k$ must be indifferent between all actions in the support of $X_k$, it must be that
    $$u_k=\pi_v(F(x))-c_k(x)=\pi_v(F(y))-c_k(y),$$ and similarly for agent of type $c_j$, it must be that
    $$u_j=\pi_v(F(x))-c_j(x)=\pi_v(F(y))-c_j(y).$$
    But this implies that $$\pi_v(F(x))-\pi_v(F(y))=c_k(x)-c_k(y)=c_j(x)-c_j(y),$$ which contradicts the fact that $\C$ is ordered.

    \item \textbf{No gaps in support:} We now show that there cannot be any gaps in the support of $X$, and that it must take the form $[0,b_K]$. Suppose instead that there is an interval $(d_1, d_2)$ which is not in the support of $X$. Then, an agent with a type that has $d_2$ in its support obtains a strictly higher payoff from choosing $d_1$, as this agent is still awarded the same expected prize, but the cost incurred by this agent is lower. It follows that the support of $X$ must be convex. An analogous argument leads to the property that the lower bound of the support must be $0$. 

    \item \textbf{Monotonicity across types:} Lastly, we show that there exist boundary points $b_1<b_2<\dots<b_K$ such that the support of $X_k$ is $[b_{k-1}, b_k]$. Suppose that $x,y$ with $x<y$ is in the support of $X_k$. We will show that for an agent of type $c_j$ where $j<k$, choosing $x$ leads to a strictly higher payoff than choosing $y$. Observe that
    $$u_k=\pi_v(F(x))-c_k(x)=\pi_v(F(y))-c_k(y).$$
    Now the payoff of agent of type $c_j$ from choosing $y$ is  
    $$\pi_v(F(y))-c_j(y)=u_k+c_k(y)-c_j(y),$$
    and that from choosing $x$ will be $$\pi_v(F(x))-c_j(x)=u_k+c_k(x)-c_j(x).$$
    Since $\C$ is ordered, $$c_j(y)-c_j(x)>c_k(y)-c_k(x) \implies c_k(x)-c_j(x) > c_k(y)-c_j(y).$$
    It follows that the agent of type $c_j$ obtains a strictly higher payoff from choosing $x$ as compared to $y$.   
\end{enumerate}

Together, the properties imply that the equilibrium exhibits the structure in the Lemma.

\end{proof}
\begin{proofsketch}
We show that a symmetric equilibrium must satisfy the following:
\begin{enumerate}
    \item The equilibrium must be in mixed strategies, and cannot have any atoms. This is because if an agent of type $c_k$ chose $x_k$ with positive probability, there is a positive probability that all agents are tied at $x_k$, and an agent of type $c_k$ would obtain a strictly higher payoff by choosing $x_k+\epsilon$ than that from choosing $x_k$ for $\epsilon>0$ and small enough.
    \item The support of the effort distribution across types should be essentially disjoint, with at most one effort level in the intersection of support of any two different types. This is because going from one effort level to another, the change in expected prize is the same irrespective of type, but the change in cost depends on the type. It follows that two different agent-types cannot both be indifferent between two different effort levels. 
    \item The supports of the different types must be connected, i.e, there shouldn't be any gaps. This is because if there is any gap $(d_1, d_2)$ in the support, an agent-type that has $d_2$ in the support would obtain a strictly higher payoff by choosing $d_1$. In doing so, the expected prize awarded to the agent remains the same, while the effort cost is lower.
    \item Finally, the effort must be monotonic in types. This is because if the distribution of type $c_k$ contains $x$ and $y$ in its support with $x<y$, then the indifference condition of type $c_k$, together with the ordered structure of $\C$, implies that for any less efficient type $c_j$ with $j<k$, choosing $x$ would lead to a strictly higher payoff than choosing $y$. 
\end{enumerate}

Together, these properties imply the result. The full proof is in the appendix.
\end{proofsketch}

Thus, for any environment $(N+1, \C, p)$ and contest $v\in \V$, the equilibrium is such that agents of the least-efficient type $c_1$ mix between $[0,b_1]$, agents of type $c_2$ mix between $[b_1, b_2]$, and so on, until we get to agents of the most-efficient agent-type $c_K$, who mix between $[b_{K-1}, b_K]$. In particular, more efficient agents always choose greater effort than less efficient agents, and the incentives to mix arise purely because of the possibility that agents might face other agents of the same type as their own. Thus, the equilibrium with finitely many types exhibits both the mixed structure characteristic of complete information environments (\citet*{barut1998symmetric}) and the monotonic structure observed in environments with a continuum of types (\citet*{moldovanu2001optimal}).  \\

It remains to identify the equilibrium distributions $(F_1, \dots, F_K)$, which can now be derived using the indifference condition. More precisely, an agent of type $c_k\in \C$ should be indifferent between all effort levels in $[b_{k-1}, b_k]$, and this uniquely pins down the equilibrium distribution $F_k$ on $[b_{k-1}, b_k]$. The following result fully characterizes the unique symmetric Bayes-Nash equilibrium of the Bayesian game. 

\begin{thm}
\label{thm:equilibrium}
Consider any contest environment $(N+1, \C, p)$ where $\C$ is an ordered type-space. For any contest $v\in \V$, the symmetric Bayes-Nash equilibrium $(X_1, \dots, X_K)$ is such that for each $k\in [K]$, the distribution $F_k:[b_{k-1}, b_k]\to [0,1]$ is defined by 
\begin{equation}
\label{eq:cdf}
    \pi_v(P_{k-1}+p_kF_k(x_k))-c_k(x_k)=u_k \text{ for all } x_k\in [b_{k-1}, b_k], 
\end{equation}
where the boundary points $b=(b_0,\dots,b_K)$, with $b_0=0$, and the equilibrium utilities $u=(u_1,\dots, u_K)$, with $u_1=0$, satisfy
\begin{equation}
\label{bounds}
\pi_v(P_{k})-c_k(b_k) = u_k \text{ for all } k \in [K],
\end{equation}
and
\begin{equation}
\label{utilities}
\pi_v(P_{k-1})-c_{k}(b_{k-1})=u_k \text{ for all } k \in [K].
\end{equation}

\end{thm}
\begin{proof}
Suppose $(X_1, X_2, \dots, X_K)$ is a symmetric Bayes-Nash equilibrium. From Lemma \ref{eqbm_intervals}, there exist boundary points $b_0<b_1< b_2< \dots< b_K$, with $b_0=0$, so that  $X_k$ is continuously distributed on $[b_{k-1}, b_k]$. It follows that an agent of type $c_k$ must be indifferent between all effort levels in this interval. Suppose $(F_1, F_2, \dots, F_K)$ is an equilibrium distribution. Notice that if an agent of type $c_k\in \C$ chooses $x_k\in [b_{k-1}, b_k]$, it outperforms any arbitrary agent with probability $P_{k-1}+p_kF_k(x_k)$, and thus, the expected value of the prize that this agent is awarded is $\pi_v(P_{k-1}+p_kF_k(x_k))$. Moreover, the cost of choosing $x_k$ is $c_k(x_k)$. Thus, by the indifference condition, the equilibrium distribution function $F_k$ must satisfy Equation \eqref{eq:cdf}.\\

It remains to solve for the boundary points and equilibrium utilities. Plugging in $x_k=b_k$ in Equation \eqref{eq:cdf} leads to Equation \eqref{bounds}, and plugging in $x_k=b_{k-1}$  in Equation \eqref{eq:cdf} leads to Equation \eqref{utilities}. Starting from $b_0=0$, Equation \eqref{utilities} gives $u_1=0$, and then, Equation \eqref{bounds} gives $b_1=c_1^{-1}(\pi_v(P_1))$. In general, once we have $b_{k-1}$, Equation \eqref{utilities} gives $u_k$, and then, Equation \eqref{bounds} gives $b_k$. Proceeding iteratively in this way, we can recover all the equilibrium boundary points and utilities from Equations \eqref{bounds} and \eqref{utilities}. Together, these three equations fully characterize the unique symmetric Bayes-Nash equilibrium of the Bayesian game.\footnote{This equilibrium characterization effectively captures equilibrium behavior in both complete and continuum type-space environments. The complete information setting where all agents have the same type $c_1\in \F$ is a special case of our model where the type-space $\C=\{c_1\}$. Moreover, by the convergence result in Appendix \ref{sec:convergence}, the (pure-strategy) equilibrium in any continuum type-space is well-approximated by the equilibrium of a sufficiently large and appropriately chosen finite type-space.}
\end{proof}

Given the equilibrium characterization, we now derive a useful representation for expected equilibrium effort. For any environment $(N+1, \C, p)$ and contest $v\in \V$, we reinterpret the equilibrium by embedding it in an alternative space--specifically, as a mapping between equilibrium effort and the probability $t \in [0,1]$ of outperforming an arbitrary agent. For instance, the boundary point $b_k$ corresponds to the equilibrium effort associated with outperforming an arbitrary agent with probability $P_k$. More generally, from Equation \eqref{eq:cdf}, the equilibrium effort associated with outperforming an arbitrary agent with probability $t \in (P_{k-1}, P_k)$ is 
$$c_k^{-1}\left(\pi_v(t)-u_{k}\right).$$

Since, ex-ante, this probability $t$ is uniformly distributed on $[0,1]$, we obtain the following representation for expected equilibrium effort.

\begin{lemrep}
\label{lem:effort}
Consider any contest environment $(N+1, \C, p)$ where $\C$ is an ordered type-space. For any contest $v\in \V$,  the expected equilibrium effort of an arbitrary agent is
$$\mathbb{E}[X]=\int_{0}^{1} g_{k(t)}\left(\pi_v(t)-u_{k(t)}\right)dt,$$
where $g_k=c_k^{-1}$ and
$k(t)=\max\{k: P_{k-1}\leq t\}$.
\end{lemrep}
\begin{proof}
We first find the expected effort exerted in equilibrium by an agent of type $c_k$. From Theorem \ref{thm:equilibrium}, we have that the (random) level of effort $X_k$ satisfies
$$\pi_v(P_{k-1}+p_kF_k(X_k))-c_k(X_k)=u_k.$$
Rearranging and taking expectations on both sides, we obtain 
\begin{align*}
    \mathbb{E}[X_k]&=\mathbb{E}\left[g_k\left(\pi_v(P_{k-1}+p_kF_k(X_k))-u_k\right)\right] &(\text{Since }g_k=c_k^{-1})\\
    &=\int_{b_{k-1}}^{b_k} g_k\left(\pi_v(P_{k-1}+p_kF_k(x_k))-u_k\right)f_k(x_k)dx_k\\
    &=\int_0^1 g_k\left(\pi_v(P_{k-1}+p_kt)-u_k\right)dt &(\text{Substituting } F_k(x_k)=t).
\end{align*}
Then,
\begin{align*}
    \mathbb{E}[X]&=\sum_{k=1}^K p_k \mathbb{E}[X_k]\\
    &=\sum_{k=1}^K p_k \int_0^1 g_k\left(\pi_v(P_{k-1}+p_kt)-u_k\right)dt\\
    &=\sum_{k=1}^K \int_{P_{k-1}}^{P_{k}} g_k\left(\pi_v(p)-u_k\right)dp &(\text{Substituting } P_{k-1}+p_kt=p)\\
    &=\int_{0}^{1} g_{k(t)}\left(\pi_v(t)-u_{k(t)}\right)dt&(\text{where }k(t)=\max\{k: P_{k-1}\leq t\})
\end{align*}
as required.
\end{proof}

This interpretation of the equilibrium as a mapping between effort and the probability of outperforming an arbitrary agent provides a unified framework for analyzing symmetric equilibrium across different environments and is central to our subsequent analysis.

\section{Effect of competition on expected effort}

In this section, we examine how increasing competitiveness of a contest influences the expected equilibrium effort, and also solve the designer's problem of allocating a fixed budget across prizes so as to maximize expected equilibrium effort.\\

From Lemma \ref{lem:effort}, it follows that for any environment $(N+1, \C, p)$ and contest $v\in \V$, the marginal effect of increasing prize $m\in [N]$ on expected effort is
\begin{equation*}
\label{eq:marginal}
\dfrac{\partial \mathbb{E}[X]}{\partial v_m}=\int_0^1 g_{k(t)}'\left(\pi_v(t)-u_{k(t)}\right)\left[H^{N}_{m}(t) -\frac{\partial u_{k(t)}}{\partial v_m}\right]dt.
\end{equation*}

And thus, for any pair of prizes $m, m'\in [N]$ with $m>m'$, the marginal effect of increasing competition by transferring value from worse-ranked prize $m'$ to better-ranked prize $m$ is  
\begin{equation}
\label{eq:competition}
\dfrac{\partial \mathbb{E}[X]}{\partial v_m} - \dfrac{\partial \mathbb{E}[X]}{\partial v_{m'}}=\int_0^1 g_{k(t)}'\left(\pi_v(t)-u_{k(t)}\right)\left[H^{N}_{m}(t)-H^{N}_{m'}(t) -\left[\frac{\partial u_{k(t)}}{\partial v_m}-\frac{\partial u_{k(t)}}{\partial v_{m'}}\right]\right]dt.
\end{equation}

To interpret Equation \eqref{eq:competition}, consider again an agent who outperforms an arbitrary agent with probability $t$. Transferring value from $m'$ to $m$ results in a marginal increase in this agent's expected prize of $H^{N}_{m}(t)-H^{N}_{m'}(t)$. By subtracting the subsequent marginal increase in utility $\left[\frac{\partial u_{k(t)}}{\partial v_m}-\frac{\partial u_{k(t)}}{\partial v_{m'}}\right]$, we isolate the marginal increase in effort costs, which is then translated into the marginal effect on effort. Finally, taking a uniform expectation over $t\in [0,1]$ gives the overall impact of the transformation on expected effort. Equation \eqref{eq:competition} provides a general and useful framework in which to think about the effect of competition on effort. We will now use this framework to analyze the effect of competition under some important contest environments.

\subsection{Complete information}

We begin with the complete information environment, captured by a type-space containing only a single type. This complete information case was the focus of \citet*{fang2020turning}, who showed that increasing competition encourages effort when the cost function is concave, and discourages effort when it is convex. We now recover this result in our framework, introducing and illustrating some key ideas that will be useful later.\\

Consider a complete information environment with type-space $\C=\{c_1\}$ where $c_1\in \F$. From Theorem \ref{thm:equilibrium}, we know that for any contest $v\in \V$, the equilibrium utility $u_1=0$. Consequently, the effect of increasing competition, as captured by Equation \eqref{eq:competition}, simplifies to
$$\dfrac{\partial \mathbb{E}[X]}{\partial v_m} - \dfrac{\partial \mathbb{E}[X]}{\partial v_{m'}}=\int_0^1 g_{1}'\left(\pi_v(t)\right)\left[H^{N}_{m}(t)-H^{N}_{m'}(t)\right]dt.$$
 
Here, observe that $\left[H^{N}_{m}(t)-H^{N}_{m'}(t)\right]$, which represents the marginal effect on effort costs, is negative for small $t$-values and positive for large $t$-values. Moreover, the aggregate effect on effort cost is $$\int_0^1 \left[H^{N}_{m}(t)-H^{N}_{m'}(t)\right]dt = 0.$$

Thus, increasing competition essentially shifts equilibrium effort costs from low $t$-values to high $t$-values. Now for the effect on effort, the term $g_1'(\pi_v(t))$ can be interpreted as assigning different weights to the effect on effort costs across different $t$-values. If these weights are monotonic in $t$ (which they are when $c_1$ is concave or convex), we can recover the effect on effort from the effect on effort costs. We formalize this idea in the following lemma.


\begin{lemrep}
\label{lem:trick}
Suppose $a_2:[0,1]\to \mathbb{R}$ is such that there exists $t^* \in [0,1]$ so that $a_2(t)\leq 0$ for $t \leq t^*$ and $a_2(t) \geq 0$ for $t \geq t^*$. Then, for any increasing function $a_1:[0,1]\to \mathbb{R}$, $$\int_0^1 a_1(t)a_2(t)dt \geq a_1(t^*)\int_0^1 a_2(t)dt.$$
\end{lemrep}
\begin{proof}
Observe that
\begin{align*}
    \int_0^1 a_1(t)a_2(t)dt&= \int_0^{t^*} a_1(t)a_2(t)dt+ \int_{t^*}^1 a_1(t)a_2(t)dt\\
    &\geq \int_0^{t^*} a_1(t^*)a_2(t)dt+ \int_{t^*}^1 a_1(t^*)a_2(t)dt\\
    &= a_1(t^*)\int_0^1 a_2(t)dt.
\end{align*}
\end{proof}

From here, a straightforward application of Lemma \ref{lem:trick} with $a_2(t)=\left[H^{N}_{m}(t)-H^{N}_{m'}(t)\right]$ leads to the following result about the effect of increasing competition on expected effort in complete information environments (\citet*{fang2020turning}).

\begin{thmrep}
\label{thm:complete}
Consider a contest environment $(N+1, \C, p)$ where $\C=\{c_1\}$ and $c_1\in \F$. For any pair $m, m'\in [N]$ with $m>m'$, the following hold:
\begin{enumerate}
    \item If $c_1$ is concave, then for any contest $v\in \V$,  $\dfrac{\partial \mathbb{E}[X]}{\partial v_m} - \dfrac{\partial \mathbb{E}[X]}{\partial v_{m'}} \geq 0$.
    \item If $c_1$ is convex, then for any contest $v\in \V$, $\dfrac{\partial \mathbb{E}[X]}{\partial v_m} - \dfrac{\partial \mathbb{E}[X]}{\partial v_{m'}} \leq 0$.
\end{enumerate}
\end{thmrep}
\begin{proof}
From Theorem \ref{thm:equilibrium}, we know that $u_1=0$, and thus, from Equation \eqref{eq:competition}, we have that 
$$\dfrac{\partial \mathbb{E}[X]}{\partial v_m} - \dfrac{\partial \mathbb{E}[X]}{\partial v_{m'}}=\int_0^1 g_{1}'\left(\pi_v(t)\right)\left[H^{N}_{m}(t)-H^{N}_{m'}(t)\right]dt.$$

If $c_1$ is concave, $g_1=c_1^{-1}$ is convex, and thus, $g_1'(\pi_v(t))$ is increasing in $t$. Applying Lemma \ref{lem:trick} with $a_1(t)=g_{1}'\left(\pi_v(t)\right)$ and $a_2(t)=\left[H^{N}_{m}(t)-H^{N}_{m'}(t)\right]$ gives the result.

If $c_1$ is convex, $g_1=c_1^{-1}$ is concave, and thus, $g_1'(\pi_v(t))$ is decreasing in $t$. Applying Lemma \ref{lem:trick} with $a_1(t)=-g_{1}'\left(\pi_v(t)\right)$ and $a_2(t)=\left[H^{N}_{m}(t)-H^{N}_{m'}(t)\right]$ gives the result.

\end{proof}

Thus, in a complete information environment, the effect of increasing competition on expected equilibrium is determined solely by the structure of the cost function. It encourages effort if the cost is concave, and discourages effort if the cost is convex. If the cost is linear, so that it is both concave and convex, increasing competition has no effect on expected effort (\citet*{barut1998symmetric}). For the design problem of allocating a budget across prizes to maximize effort, the solution follows directly from Theorem \ref{thm:complete}, and we note it in the following corollary.

\begin{corollary}
\label{cor:complete}
Consider a contest environment $(N+1, \C, p)$ where $\C=\{c_1\}$ and $c_1\in \F$. Suppose any contest $v\in \V$ such that $\sum_{m=0}^N v_m\leq V$ is feasible. 
\begin{enumerate}
    \item If $c_1$ is strictly concave, the contest $v=(0, 0, \dots, 0, V)$ uniquely maximizes $\E[X]$.
    \item If $c_1$ is linear, any contest $v\in \V$ such that $\sum_{m=1}^{N} v_m=V$ maximizes $\E[X]$.
    \item If $c_1$ is strictly convex, the contest $v=\left(0, \frac{V}{N}, \dots, \frac{V}{N}\right)$ uniquely maximizes $\E[X]$.
\end{enumerate}
\end{corollary}

\subsection{Incomplete information: Linear cost}
\label{subsec:inlinear}
We now turn to the incomplete information environment. Compared to the complete information case, the analysis here is more nuanced, as increasing competition not only effects the equilibrium effort, but also the equilibrium utilities of the different agent-types. Moreover, these effects on utilities may depend in an intricate way on the specific structure of the type-space $\C$. To begin, we focus on the special case where all cost functions are linear.\\

Consider a contest environment $(N+1, \C, p)$ where $\C$ is such that $c_k(x)=\th_k\cdot x$, with $\th_1>\dots>\th_K>0$. In this case, it turns out that for any contest $v\in \V$, we can explicitly solve for the expected effort. Using Lemma \ref{lem:effort}, we can express the expected effort as
\begin{align*}
\mathbb{E}[X]&=\sum_{k=1}^K p_k \cdot \frac{1}{\th_k}\cdot\left[\int_{P_{k-1}}^{P_k}  \frac{\pi_v(t)}{p_k}dt-u_k\right],
\end{align*}
Here, $\displaystyle\int_{P_{k-1}}^{P_k} \dfrac{\pi_v(t)}{p_k}dt$ is simply the expected prize awarded to an agent of type $c_k\in \C$, and is linear in $v_m$ for $m\in [N]$. Further, using Equations \eqref{bounds} and \eqref{utilities}, we can solve for the equilibrium utilities and show that:
\begin{equation}
\label{eq:utilities}
u_k=\theta_k \left[\sum_{j=1}^{k-1} \pi_v(P_{j}) \left(\frac{1}{\theta_{j+1}}-\frac{1}{\theta_j}\right) \right] \text{ for  } k\in [K],
\end{equation}
which is also linear in $v_m$ for $m\in [N]$. Substituting these expressions, we derive the following representation for the expected equilibrium effort.

\begin{lemrep}
\label{lem:linear}
Consider a contest environment $(N+1, \C, p)$ where $\C$ is such that $c_k(x)=\th_k\cdot x$, with $\th_1>\dots>\th_K>0$. For any contest $v\in \V$, the expected equilibrium effort is $$\mathbb{E}[X]=\sum_{m=1}^N \alpha_m v_m,$$
where
\begin{equation}
\label{eq:linear_effort}
\alpha_m=\frac{1}{N+1}\left[\frac{1}{\theta_K}-\sum_{k=1}^{K-1} \left[H^{N+1}_{\geq m}(P_k)+(N-m)H^{N+1}_{m}(P_k)\right]\left(\frac{1}{\theta_{k+1}}-\frac{1}{\theta_{k}}\right)\right].
\end{equation}
\end{lemrep}
\begin{proof}
Using the representation in Lemma \ref{lem:effort}, we have that for any contest $v\in \V$, 
\begin{align*}
\mathbb{E}[X]&=\int_{0}^{1} g_{k(t)}\left(\pi_v(t)-u_{k(t)}\right)dt\\
&=\int_{0}^{1} \dfrac{\left(\pi_v(t)-u_{k(t)}\right)}{\th_{k(t)}}dt &\left(g_k(y)=\frac{y}{\th_k}\right)\\
&=\sum_{k=1}^K p_k \cdot \frac{1}{\th_k}\cdot\left[\int_{P_{k-1}}^{P_k}  \frac{\pi_v(t)}{p_k}dt-u_k\right].
\end{align*}

\begin{enumerate}
    \item Notice that for any $k\in [K]$,  $\displaystyle\int_{P_{k-1}}^{P_k} \frac{\pi_v(t)}{p_k}dt$ is the expected prize awarded to an agent of type $c_k$. To compute this, we instead compute the ex-ante expected total prize awarded to agents of type $c_k$. Notice that for any prize $m\in \{0, \dots,  N\}$, the ex-ante probability that this prize is awarded to an agent of type $c_k$ is simply $$\left[H^{N+1}_{\geq m+1}(P_k)-H^{N+1}_{\geq m+1}(P_{k-1})\right].$$  
    Thus, the ex-ante expected total prize awarded to agents of type $c_k$ is 
     $$\sum_{m=1}^N v_m \left[H^{N+1}_{\geq m+1}(P_k)-H^{N+1}_{\geq m+1}(P_{k-1})\right].$$ 

     By an alternative calculation, which entails adding up over the $N+1$ agents, this expectation should equal $$(N+1)\cdot p_k \cdot \displaystyle\int_{P_{k-1}}^{P_k} \frac{\pi_v(t)}{p_k}dt.$$

     Equating these two, we get that
     $$\int_{P_{k-1}}^{P_k} \pi_v(t)dt=\dfrac{\sum_{m=1}^N v_m \left[H^{N+1}_{\geq m+1}(P_k)-H^{N+1}_{\geq m+1}(P_{k-1})\right]}{N+1}.$$
Alternatively, we can also directly use the following fact to compute this integral:

$$\dfrac{\partial H^{N+1}_{\geq m+1}(t)}{\partial t} = (N+1)H^{N}_{m}(t)$$
\item For the equilibrium utilities $u_k$, we simply solve the Equations \eqref{bounds} and \eqref{utilities}. For the given type-space $\C$ with $c_k(x)=\th_k\cdot x$, these equations can be rewritten as
$$\pi_v(P_k)-\th_k b_k = u_k \text{ and } \pi_v(P_{k-1})-\th_kb_{k-1} = u_k.$$
Solving this system of equations gives
$$
b_k=\sum_{j=1}^k \dfrac{\pi_v(P_j)-\pi_v(P_{j-1})}{\theta_j} \text{ for } k\in [K],
$$
and
$$
u_k=\theta_k \left[\sum_{j=1}^{k-1} \pi_v(P_{j}) \left(\frac{1}{\theta_{j+1}}-\frac{1}{\theta_j}\right) \right] \text{ for } k\in [K].
$$

\end{enumerate}

Substituting these expressions in the above representation, we get that
$$\mathbb{E}[X] = \sum_{k=1}^K \frac{1}{(N+1)\th_k}\sum_{m=1}^N v_m \left[H^{N+1}_{\geq m+1}(P_k)-H^{N+1}_{\geq m+1}(P_{k-1})\right]-\sum_{k=1}^K \dfrac{p_ku_k}{\th_k}.$$

From here, it follows that we can write $$\E[X]=\sum_{m=1}^N \alpha_mv_m$$ where
\begin{align*}
    \alpha_m&= \sum_{k=1}^K \frac{\left[H^{N+1}_{\geq m+1}(P_k)-H^{N+1}_{\geq m+1}(P_{k-1})\right]}{(N+1)\th_k} - \sum_{k=1}^K p_k \sum_{j=1}^{k-1} H^N_{m}(P_j)\left(\frac{1}{\theta_{j+1}}-\frac{1}{\theta_j}\right)\\
    &= \sum_{k=1}^K \frac{\left[H^{N+1}_{\geq m+1}(P_k)-H^{N+1}_{\geq m+1}(P_{k-1})\right]}{(N+1)\th_k} - \sum_{k=1}^{K-1} (1-P_k)H^N_{m}(P_k)\left(\frac{1}{\theta_{k+1}}-\frac{1}{\theta_k}\right)\\
    &=\frac{1}{N+1}\left[\frac{1}{\theta_K}-\sum_{k=1}^{K-1} H^{N+1}_{\geq m+1}(P_k)\left(\frac{1}{\theta_{k+1}}-\frac{1}{\theta_{k}}\right)\right] - \frac{(N+1-m)}{N+1}\sum_{k=1}^{K-1} \left[H^{N+1}_{m}(P_k)\left(\frac{1}{\theta_{k+1}}-\frac{1}{\theta_{k}}\right)\right]\\
    &=\frac{1}{N+1}\left[\frac{1}{\theta_K}-\sum_{k=1}^{K-1} \left[H^{N+1}_{\geq m}(P_k)+(N-m)H^{N+1}_{m}(P_k)\right]\left(\frac{1}{\theta_{k+1}}-\frac{1}{\theta_{k}}\right)\right].
\end{align*}
\end{proof}
Thus, in the incomplete information environment with linear types, the expected equilibrium effort is linear in the values of the different prizes, with coefficients that depend on the specifics of the environment. It follows then that the effect of increasing competition in this environment, as captured by Equation \eqref{eq:competition}, simplifies to 
$$\dfrac{\partial \mathbb{E}[X]}{\partial v_m} - \dfrac{\partial \mathbb{E}[X]}{\partial v_{m'}}=\alpha_m-\alpha_{m'},$$
which can now be explicitly evaluated using Equation \eqref{eq:linear_effort}.\\

In particular, we first note that increasing competition by transferring value to the best-ranked prize always encourages effort. To see why, notice from Equation \eqref{eq:linear_effort} that for any prize $m'\in \{1, \dots, N-1\}$, 
$$\alpha_N-\alpha_{m'}=\frac{1}{N+1}\left[\sum_{k=1}^{K-1} \left[H^{N+1}_{\geq {m'}}(P_k)-H^{N+1}_{\geq N}(P_k)+(N-m')H^{N+1}_{m'}(P_k)\right]\left(\frac{1}{\theta_{k+1}}-\frac{1}{\theta_{k}}\right)\right].$$
With $m'<N$ and $K\geq 2$, it is straightforward to verify that $\alpha_N-\alpha_{m'}>0$. In other words, for any contest $v\in \V$, transferring value from any lower-ranked prize $m'$ to the top-prize $N$ leads to an increase in expected effort. Consequently, for the design problem, it follows that allocating the entire budget to the best-performing agent is strictly optimal. 
\begin{corollary}
\label{cor:linear}
Consider a contest environment $(N+1, \C, p)$ where $\C$ is such that $c_k(x)=\th_k\cdot x$, with $\th_1>\dots>\th_K>0$ and $K\geq 2$. Among all contests $v\in \V$ such that $\sum_{m=0}^N v_m\leq V$, the contest $v=(0, 0, \dots,  0, V)$ uniquely maximizes $\E[X]$.    
\end{corollary}

This result resolves the conjecture of \citet*{sisak2009multiple}, which suggested that allocating the budget across multiple prizes might be optimal in this environment. Additionally, it extends the optimality of the winner-takes-all contest under linear costs, previously established for a continuum type-space environment by \citet*{moldovanu2001optimal}, to the finite type-space setting. Thus, as soon as there is any uncertainty (incomplete information) in an environment with linear costs, the winner-takes-all contest is strictly optimal.\\

Even though transferring value to the best-ranked prize always encourages effort, increasing competition by transferring value to better-ranked intermediate prizes may not always encourage effort. To see this, consider a contest environment with just two types. For this case with $K=2$, we can show that for any $m\in \{1, \dots, N-1\}$, the marginal effect of transferring value from prize $m$ to prize $m+1$ on expected effort is
\begin{align*}
\alpha_{m+1}-\alpha_{m}\geq 0 &\iff P_1\leq \frac{m+1}{N}.
\end{align*}
It follows that the effect of the transformation depends on the relative likelihood of efficient and inefficient types. In particular, if $P_1$ is small ($P_1< \frac{2}{N}$), increasing competition by transferring value to better-ranked prizes always encourages effort. However, if $P_1$ is big ($P_1>\frac{N-1}{N}$), increasing competition actually generally discourages effort, except when transferring value to the best-ranked prize. Intuitively, increasing competition encourages effort from the efficient types, while discouraging effort from the inefficient types. Thus, if the population is more likely to be efficient, the overall effect is positive, but if it is more likely to be inefficient, increasing competition by transferring value across intermediate prizes may actually discourage effort.

\subsection{Incomplete information: General cost}

In this subsection, we continue our analysis of the incomplete information environment, allowing for more general cost functions. Unlike the case of linear costs, where expected effort depends linearly on the value of prizes, the precise relationship between expected effort and the prize values under general costs may be complex. For tractability, we restrict our attention to parametric type-spaces. 
\\

Consider a contest environment $(N+1, \C, p)$ where $\C$ is a parametric type-space, defined by parameters $\th_1>\dots >\th_K$ and a (base) cost function $c\in \F$, so that $c_k(x)=\th_k\cdot c(x)$. In this case, if we let $g=c^{-1}$, notice that we can express $g_k(y)=g\left(\frac{y}{\th_k}\right)$, so that $g_k'(y)=\frac{1}{\th_k}g'\left(\frac{y}{\th_k}\right)$. Thus, for any contest $v\in \V$, the effect of increasing competition on expected effort, as captured by Equation \eqref{eq:competition}, can be expressed as
\begin{align}
\dfrac{\partial \mathbb{E}[X]}{\partial v_m} - \dfrac{\partial \mathbb{E}[X]}{\partial v_{m'}}&=\int_0^1 g'\left(\frac{\pi_v(t)-u_{k(t)}}{\th_{k(t)}}\right)\left[\frac{H^{N}_{m}(t)-H^{N}_{m'}(t)}{\th_{k(t)}} -\frac{1}{\th_{k(t)}}\left[\frac{\partial u_{k(t)}}{\partial v_m}-\frac{\partial u_{k(t)}}{\partial v_{m'}}\right]\right]dt \notag\\
&=\int_0^1 g'\left(\frac{\pi_v(t)-u_{k(t)}}{\th_{k(t)}}\right)\left[\lambda_m(t)-\lambda_{m'}(t)\right]dt, \label{eq:competition_para}
\end{align}
where 
$$\lambda_m(t)=\left(\frac{H^{N}_{m}(t)}{\theta_{k(t)}} -\frac{1}{\theta_{k(t)}}\frac{\partial u_{k(t)}}{\partial v_m}\right).$$
To analyze this, we first discuss how increasing competition affects the equilibrium utilities of the different agent types. Fix any cost function $c\in \F$ and any contest $v\in \V$. Notice that we can reinterpret the induced Bayesian game as one where agents directly choose effort cost, $c(x)$, instead of choosing effort $x$. Consequently, the properties of equilibrium effort $x$ under linear costs, derived in Subsection \ref{subsec:inlinear}, actually more generally represent properties of equilibrium effort cost $c(x)$ under cost function $c$.\footnote{As a corollary, for any contest environment $(N+1, \C, p)$ where $\C$ is a parametric type-space with base function $c\in \F$, the winner-takes-all contest $v=(0, \dots, 0, V)$ uniquely maximizes $\E[c(X)]$ among all contests feasible with a budget $V$.} In particular, it follows that the equilibrium utilities are exactly as those described in Equation \eqref{eq:utilities}, so that the marginal effect of increasing competition on utility of type $c_k(\cdot) = \th_k \cdot c(\cdot)$ is
$$
\dfrac{\partial u_k}{\partial v_m} - \dfrac{\partial u_k}{\partial v_{m'}}=\th_k\left[\sum_{j=1}^{k-1} H^{N}_{m}(P_j) - H^{N}_{m'}(P_j)  \left(\frac{1}{\theta_{j+1}}-\frac{1}{\theta_j}\right) \right].
$$
Notice that this effect of competition on equilibrium utilities is independent of the cost function $c\in \F$, and also the contest $v \in \V$.\\

We now return to analyzing the effect of competition on equilibrium effort. In Equation \eqref{eq:competition_para}, it follows that the cost function $c\in \F$ only influences the first term of the integrand, $g'\left(\frac{\pi_v(t)-u_{k(t)}}{\th_{k(t)}}\right)$. As in our analysis of the complete information case, we interpret this term as simply assigning different weights across different $t$-values. The second term, $\lambda_m(t)-\lambda_{m'}(t)$, captures the marginal effect of the transformation on (base) effort costs across different $t$-values. From our analysis of the linear cost case $(c(x)=x)$, we know that
$$\int_0^1 \left[\lambda_m(t)-\lambda_{m'}(t)\right]dt = \alpha_m-\alpha_{m'},$$
which we can now interpret more generally as the marginal effect of increasing competition on expected equilibrium (base) effort cost. If this second term, $\lambda_m(t)-\lambda_{m'}(t)$, further exhibits the single-crossing property, as it does in the complete information case, this effect on effort cost may be informative about the effect on effort itself. It is straightforward to verify that the term $\lambda_m(t)-\lambda_{m'}(t)$ is continuous in $t$, and moreover, has the same sign of the derivative as $H^N_m(t)-H^N_{m'}(t)$, whenever it exists. It follows that the term is $0$ at $t=0$, initially decreases, then increases, and eventually decreases again (unless $m=N$). Thus, $\lambda_m(1)-\lambda_{m'}(1)\geq 0$ is both necessary and sufficient to ensure that $\lambda_m(t)-\lambda_{m'}(t)$ is single-crossing in $t$. In other words, while increasing competition reduces the effort cost associated with low $t$-values, the condition ensures that it leads to an increase in effort cost associated with all the larger $t$-values.  \\

With this, we are now ready to state our main result. The result identifies conditions under which $\lambda_m(1)-\lambda_{m'}(1)\geq 0$ and establishes how, under these conditions, the effect of increasing competition on expected equilibrium effort under general costs may be inferred from its effect on effort costs (or alternatively, from its effect on effort under linear costs).

\begin{thmrep}
\label{thm:diff_general}
Consider a contest environment $(N+1, \C, p)$ where $\C$ is a parametric type-space, defined by parameters $\th_1>\dots >\th_K$ and (base) function $c\in \F$, so that $c_k(x)=\th_k\cdot c(x)$. Let $m, m' \in [N]$ with $m>m'$ be such that, either $m=N$ or $$\left(\dfrac{\partial u_K}{\partial v_m} - \dfrac{\partial u_K}{\partial v_{m'}}\right) \leq 0 \iff \sum_{k=1}^{K-1} \left(H^{N}_{m}(P_k)- H^{N}_{m'}(P_k)\right) \left(\frac{1}{\theta_{k+1}}-\frac{1}{\theta_k}\right)
\leq 0.$$ 
Then, the following hold:
\begin{enumerate}
    \item If $\alpha_m-\alpha_{m'}\geq 0$ and $c$ is concave,  then for any contest $v \in \V$, $\dfrac{\partial \mathbb{E}[X]}{\partial v_m} - \dfrac{\partial \mathbb{E}[X]}{\partial v_{m'}}\geq  0.$
    \item If $\alpha_m-\alpha_{m'}\leq 0$ and $c$ is convex, then for any contest $v \in \V$, $\dfrac{\partial \mathbb{E}[X]}{\partial v_m} - \dfrac{\partial \mathbb{E}[X]}{\partial v_{m'}}\leq 0$.
\end{enumerate}
\end{thmrep}
\begin{proof}

For the given parametric type-space, we have from Equation \eqref{eq:competition} that for any contest $v\in \V$ and any pair of prizes $m, m' \in [N]$ with $m>m'$, 
$$
\dfrac{\partial \mathbb{E}[X]}{\partial v_m} - \dfrac{\partial \mathbb{E}[X]}{\partial v_{m'}}=\int_0^1 g'\left(\frac{\pi_v(t)-u_{k(t)}}{\th_{k(t)}}\right)\left(\lambda_m(t)-\lambda_{m'}(t)\right)dt,
$$
where 
$$\lambda_m(t)=\left(\frac{H^{N}_{m}(t)}{\theta_{k(t)}} -\frac{1}{\theta_{k(t)}}\frac{\partial u_{k(t)}}{\partial v_m}\right).$$

Further, we know from Theorem \ref{thm:equilibrium} that the equilibrium boundary points $b=(b_1, \dots, b_K)$ and utilities $u=(u_1, \dots, u_K)$ must satisfy Equations \eqref{bounds} and \eqref{utilities}. Solving these equations, we get that the equilibrium utilities are as described in Equation \eqref{eq:utilities}, and thus, we get that 
$$
\dfrac{\partial u_k}{\partial v_m}=\th_k\left[\sum_{j=1}^{k-1} H^{N}_{m}(P_j) \left(\frac{1}{\theta_{j+1}}-\frac{1}{\theta_j}\right) \right].
$$

Plugging in, we get that
\begin{align*}
\lambda_m(t)-\lambda_{m'}(t)=\left(\frac{H^{N}_{m}(t) - H^{N}_{m'}(t)}{\theta_{k(t)}}\right) -\left[\sum_{j=1}^{k(t)-1} \left(H^{N}_{m}(P_j) -H^{N}_{m'}(P_j)\right) \left(\frac{1}{\theta_{j+1}}-\frac{1}{\theta_j}\right) \right].
\end{align*}
From here, one can verify that
\begin{enumerate}
    \item $\lambda_m(0)-\lambda_{m'}(0)=0$
    \item $\lambda_m(1)-\lambda_{m'}(1)=\begin{cases}\frac{1}{\theta_K}-\frac{1}{\theta_K}\left(\frac{\partial u_K}{\partial v_m} - \frac{\partial u_K}{\partial v_{m'}}\right)  & \text { if } m=N \\ -\frac{1}{\theta_K}\left(\frac{\partial u_K}{\partial v_m} - \frac{\partial u_K}{\partial v_{m'}}\right) & \text { otherwise } \end{cases}$
    \item $\lambda_m(t)-\lambda_{m'}(t)$ is continuous in $t$
    \item $\lambda_m(t)-\lambda_{m'}(t)$ is differentiable at $t\in [0,1]$ for $t\neq P_k$, and at any such $t$, the derivative has the same sign as the derivative of $H^{N}_{m}(t) - H^{N}_{m'}(t)$ with respect to $t$.
\end{enumerate}

Since $m, m'$ are such that either $m=N$ or  $\left(\dfrac{\partial u_K}{\partial v_m} - \dfrac{\partial u_K}{\partial v_{m'}}\right)\leq 0$, we get that $\lambda_m(1)-\lambda_{m'}(1)\geq 0$. Together with the above properties, this implies that there is some $t^*\in [0,1]$ such that $\lambda_m(t)-\lambda_{m'}(t)\leq 0$ for $t\in [0, t^*]$, and $\lambda_m(t)-\lambda_{m'}(t)\geq 0$ for $t\in [t^*, 1]$. \\

Now if $c$ is concave, $g=c^{-1}$ is convex, and thus, $g'\left(\frac{\pi_v(t)-u_{k(t)}}{\th_{k(t)}}\right)$ is increasing in $t$. Applying Lemma \ref{lem:trick} with $a_1(t)=g'\left(\frac{\pi_v(t)-u_{k(t)}}{\th_{k(t)}}\right)$ and $a_2(t)=\lambda_m(t)-\lambda_{m'}(t)$ gives 
\begin{align*}
\dfrac{\partial \mathbb{E}[X]}{\partial v_m} - \dfrac{\partial \mathbb{E}[X]}{\partial v_{m'}} &\geq g'\left(\frac{\pi_v(t^*)-u_{k(t^*)}}{\th_{k(t^*)}}\right) \int_0^1 (\lambda_m(t)-\lambda_{m'}(t))dt\\
&=g'\left(\frac{\pi_v(t^*)-u_{k(t^*)}}{\th_{k(t^*)}}\right) \left(\alpha_m-\alpha_{m'}\right)
\end{align*}
and the result follows. An analogous argument applies for the case where $c$ is convex.

\end{proof}

In words, if increasing competition does not lead to an increase in the equilibrium utility of the most-efficient type, or if it involves transferring value to the best-ranked prize, then its effect on effort may be inferred from its effect on effort costs. In such cases, for any contest $v\in \V$, the effect on effort costs (or effort under linear costs) extends to the effect on effort under concave costs if it is positive, and to the effect on effort under convex costs if it is negative. Despite being somewhat limited in its scope, Theorem \ref{thm:diff_general} provides a convenient method to check if increasing competitiveness of a contest would encourage or discourage effort under fairly general environments. \\

In particular, it allows us to solve the design problem of allocating a budget across prizes for the case of concave costs. To see how, fix any contest environment $(N+1, \C, p)$ where $\C$ is a parametric type-space with a concave cost $c\in \F$. For any contest $v\in \V$, consider the effect of transferring value from an arbitrary prize $m' \in \{1, \dots, N-1\}$ to the best-ranked prize $m=N$. From Theorem \ref{thm:diff_general}, if $\alpha_N-\alpha_{m'}\geq 0$, the transformation will have an encouraging effect on expected equilibrium effort. But we know from our analysis of the linear costs in Subsection \ref{subsec:inlinear} that $\alpha_N-\alpha_{m'}\geq 0$. It follows that for any concave cost $c\in \F$ and contest $v\in \V$, transferring value to the best-ranked prize encourages expected equilibrium effort. As a result, it is optimal to allocate the entire budget to the top-ranked prize.

\begin{corollary}
\label{cor:concave}
Consider a contest environment $(N+1, \C, p)$ where $\C$ is a parametric type-space, defined by parameters $\th_1>\dots >\th_K$ and (base) function $c\in \F$, so that $c_k(x)=\th_k\cdot c(x)$. If $c$ is concave, among all contests $v\in \V$ such that $\sum_{m=0}^N v_m\leq V$, the contest $v=(0, 0, \dots,  0, V)$ uniquely maximizes $\E[X]$.    
\end{corollary}

This result extends the optimality of the winner-takes-all contest under concave costs, previously established for a continuum type-space environment by \citet*{moldovanu2001optimal}, to the finite type-space setting. Together with Corollaries \ref{cor:complete} and \ref{cor:linear}, it follows that the winner-takes-all contest is robustly optimal under concave costs, irrespective of whether the environment is a complete or incomplete information environment.

\begin{toappendix}
\section{Convergence to continuum type-space equilibrium}
\label{sec:convergence}
In this section, we show that for any (parametric) continuum type-space and differentiable distribution over this type-space, if we take a sequence of (parametric) finite type-space distributions that converge to this distribution, the corresponding sequence of mixed-strategy equilibrium converges to the pure-strategy equilibrium under the continuum type-space. Intuitively, as the finite type-space becomes large, the interval over which an agent of a certain type mixes shrinks, and essentially converges to the effort level prescribed by the pure-strategy equilibrium under the continuum type-space. Thus, the equilibrium strategy in an appropriate and sufficiently large finite-type space domain provides a reasonable approximation to the equilibrium strategy under the continuum type-space, and vice versa.\\

We restrict attention to parametric type-spaces with linear costs ($c(x)=x$), and note that this is without loss of generality because of the equivalence between convergence properties of equilibrium costs and equilibrium effort. First, we note the symmetric equilibrium under a (parametric) continuum type-space (\citet{moldovanu2001optimal}).

\begin{lemrep}
\label{moldovanu}
Suppose there are $N+1$ agents, each with a private type (marginal cost of effort) drawn from $\Theta=[\underline{\theta}, \overline{\theta}]$ according to a differentiable CDF $G:[\underline{\theta}, \overline{\theta}]\to [0, 1]$. For any contest $v\in \V$, there is a unique symmetric Bayes-Nash equilibrium and it is such that for any $\theta\in \Theta$,
$$X(\theta)=\int_\theta^{\overline{\theta}} \dfrac{\pi_v'(1-G(t))g(t)}{t}dt.$$
\end{lemrep}
\begin{proof}
Suppose $N$ agents are playing a strategy $X: [\underline{\theta}, \overline{\theta}] \to \mathbb{R}_+$ so that if an agent's type is $\theta$, it exerts effort $X(\theta)$. Further suppose that $X(\theta)$ is decreasing in $\theta$. Now we want to find the remaining agent's best response to this strategy of the other agents. If the agent's type is $\theta$ and it pretends to be an agent of type $t \in [\underline{\theta}, \overline{\theta}]$, its payoff is
$$\pi_v(1-G(t))-\theta X(t).$$ 
Taking the first order condition, we get
$$\pi_v'(1-G(t))(-g(t))-\theta X'(t)=0.$$
Now we can plug in $t=\theta$ to get the condition for $X(\theta)$ to be a symmetric Bayes-Nash equilibrium. Doing so, we get 
$$\pi_v'(1-G(\theta))(-g(\theta))-\theta X'(\theta)=0$$ so that
$$X(\theta)=\int_\theta^{\overline{\theta}} \dfrac{\pi_v'(1-G(t))g(t)}{t}dt.$$
\end{proof}

We now state and prove the convergence result.

\begin{thmrep}
\label{thm:convergence}
Suppose there are $N+1$ agents and consider any contest $v\in \V$. Let $G:[\underline{\theta},\overline{\theta}]\to [0, 1]$ be a differentiable CDF and let $G^1, G^2, \dots, $ be any sequence of CDF's, each with a finite support, such that for all $\theta \in [\underline{\theta},\overline{\theta}]$,   $$\lim_{n \to \infty} G^n(\theta) = G(\theta).$$
Let $F^n:\mathbb{R}\to [0,1]$ denote CDF of the equilibrium effort under the finite type-space distribution $G^n$, and let $F:\mathbb{R}\to [0,1]$ denote CDF of the equilibrium under continuum type-space distribution $G$. Then, the sequence of CDF's $F^1, F^2, \dots, $ converges to the CDF $F$, i.e., for all $x\in \mathbb{R}$, 
$$\lim_{n \to \infty} F^n(x) = F(x).$$
\end{thmrep}

\begin{proof}
For the finite support CDF $G^n$, let $\Theta^n = (\theta^n_1, \theta^n_2, \dots, \theta^n_{K(n)})$ denote the support and $p^n = (p^n_1, p^n_2, \dots, p^n_{K(n)})$ denote the probability mass function. From Theorem \ref{thm:equilibrium}, let $b^n=(b^n_1, b^n_2, \dots, b^n_{K(n)})$ denote the boundary points, $u^n=(u^n_1, u^n_2, \dots, u^n_{K(n)})$ denote the equilibrium utilities, and $F^n_k$ denote the equilibrium CDF of agent of type $\theta^n_k$ on support $[b^n_{k-1}, b^n_k]$. Then, the CDF of the equilibrium effort, $F^n:\mathbb{R}\to [0,1]$, is such that for any $x\in \mathbb{R}$, 
\begin{equation}
\label{eq:cdffinite}
F^n(x)= \begin{cases} 0 & \text{if } x\leq 0\\
P^n_{k-1}+p^n_kF^n_k(x) & \text{if } x\in [b^n_{k-1}, b^n_k] \\ 1 & \text{if } x\geq b^n_{K(n)} \\ \end{cases}.    
\end{equation}

For the continuum CDF $G:[\underline{\theta}, \overline{\theta}]\to [0, 1]$,  the CDF of the equilibrium effort, $F:\mathbb{R}\to [0,1]$, is such that for any $x\in \mathbb{R}$, 

\begin{equation}
\label{eq:cdfcontinuum}
F(x)= \begin{cases} 0 & \text{if } x\leq 0\\
1-G(\theta(x)) & \text{if } x\in [0, B] \\ 1 & \text{if } x\geq B \\ \end{cases}.    
\end{equation}
where $\theta(x)$ is the inverse of $X(\theta)$ (from Lemma \ref{moldovanu}) and $B=X(\underline{\theta})$. The following Lemma will be the key to showing that $F^n(x)$ converges to $F(x)$ for all $x\in \mathbb{R}$.

\begin{lem}
\label{bconverge}
Consider any $\theta\in (\underline{\theta}, \overline{\theta})$ and for any $n\in \mathbb{N}$, let $k(n) \in \{0, 1, 2, \dots, K(n)\}$ be such that $\theta^n_{k(n)} > \theta \geq \theta^n_{k(n)+1}$ (where $\theta^n_0=\infty$ and $\theta^n_{K(n)+1}=0$). 
Then, $$\lim_{n\to \infty} b^n_{k(n)} = X(\theta) \text{ and } \lim_{n\to \infty} F^n(b^n_{k(n)}) = 1-G(\theta).$$
\end{lem}
\begin{nestedproof}
From Lemma \ref{moldovanu} and Theorem \ref{thm:equilibrium}, we have  $$X(\theta) = \int_\theta^{\overline{\theta}} \dfrac{\pi_v'(1-G(t))g(t)}{t}dt \text{ and } b^n_{k(n)}=\sum_{j=1}^{k(n)} \dfrac{\pi_v(P^n_j)-\pi_v(P^n_{j-1})}{\theta^n_j}.$$ 
Observe that
\begin{align*}
b^n_{k(n)}&=\left[\frac{\pi_v(P^n_{k(n)})}{\theta^n_{k(n)}}-\sum_{j=1}^{k(n)-1} \pi_v(P^n_j)\left[\frac{1}{\theta^n_{j+1}}-\frac{1}{\theta^n_{j}}\right]\right]\\
&=\int_0^{1/\theta^n_{k(n)}} \left[\pi_v(P^n_{k(n)}) - \pi_v(1-G^n(1/x)) \right]dx \\
& \xrightarrow{n \rightarrow \infty} \int_0^{\frac{1}{\theta}} \left[\pi_v(1-G(\theta))-\pi_v(1-G(1 / x))\right] dx \quad \text { (dominated convergence) }\\
& =\underbrace{[x(\pi_v(1-G(\theta))-\pi_v(1-G(1 / x)))]_0^{\frac{1}{\theta}}}_{\text {this is } 0}+\int_0^{\frac{1}{\theta}} \frac{\pi_v'(1-G(1 / x)) g(1 / x)}{x} dx\\
&=\int_{\theta}^{\infty} \frac{\pi_v'(1-G(t)) g(t)}{t} \mathrm{~d} t \quad \text { (substitute } t=1 / x)\\
&=X(\theta)
\end{align*}

By definition, we have
\begin{align*}
\lim_{n\to \infty} F^n(b^n_{k(n)}) &= \lim_{n\to \infty} P^n_{k(n)}\\
&= \lim_{n\to \infty} \left[1-G^n(\theta)\right]\\
&=1-G(\theta)
\end{align*}
\end{nestedproof}

Returning to the proof of Theorem \ref{thm:convergence}, fix any $x\in (0,B)$ and let $\theta \in (\underline{\theta}, \overline{\theta})$ be such that $X(\theta)=x$. Then, we know that  $$F(x)=1-G(\theta).$$ 
We want to show that  $$\lim_{n \to \infty} F^n(x) = 1-G(\theta).$$
Take $\epsilon>0$. Find $\theta'<\theta$ and $\theta''>\theta$ such that $$0<G(\theta)-G(\theta') = G(\theta'')-G(\theta)<\frac{\epsilon}{4}.$$
Let $x'=X(\theta')$, $x''=X(\theta'')$, so that $x'>x>x''$. Let $\delta = \min\{x'-x, x-x''\}$. From Lemma \ref{bconverge},  let $N(\epsilon)$ be such that for all $n>N(\epsilon)$, 
$$\max\{|b^n_{k(n)}-x|, |b^n_{k'(n)} - x'|, |b^n_{k''(n)} - x''|\} <\frac{\delta}{2}$$ and  $$\max\{|F^n(b^n_{k'(n)})-(1-G(\theta'))|, |F^n(b^n_{k''(n)})-(1-G(\theta''))|\} <\frac{\epsilon}{4},$$
where $k(n), k'(n), k''(n)$ are sequences as defined in Lemma \ref{bconverge} for $\theta$, $\theta'$ and $\theta''$ respectively. Then, for all $n>N(\epsilon)$,
\begin{align*}
    F^n(x)& > F^n(b^n_{k''(n)})\\
    &> 1-G(\theta'')-\frac{\epsilon}{4}\\
    &> 1-G(\theta)-\frac{\epsilon}{2}
\end{align*}
and 
\begin{align*}
    F^n(x)& < F^n(b^n_{k'(n)})\\
    &< 1-G(\theta')+\frac{\epsilon}{4}\\
    &< 1-G(\theta)+\frac{\epsilon}{2}
\end{align*}
so that $|F^n(x)-(1-G(\theta))|<\epsilon$. Thus, $\lim_{n \to \infty} F^n(x)=1-G(\theta)=F(x)$ for all $x\in \mathbb{R}$.
\end{proof}


\end{toappendix} 
\section{Conclusion}

This paper studies the effect of increasing competitiveness of contests in a finite type-space environment. First, we completely characterize the unique symmetric Bayes-Nash equilibrium, showing that it entails different agent types mixing over disjoint but connected intervals, so that more efficient agents exert greater effort than less efficient ones. We reinterpret the equilibrium as a mapping between effort and the probability of outperforming an arbitrary agent, which provides a unifying framework for studying contests across different environments. We then show that the effect of increasing competition on effort is more nuanced than in the complete information case because of the additional effect it has on the equilibrium utilities of the more efficient types.  We explicitly solve for these effects under linear costs and identify conditions under which they extend to general costs. Our findings suggest that increasing competition encourages effort if efficient types are likely, while discouraging effort if they're unlikely. We derive implications for the classical design problem of allocating a budget across prizes, establishing the winner-takes-all contest as being robustly optimal under linear and concave costs.\\

We hope that our results and methods will encourage further research in this fundamental finite type-space domain. The question of how competition effects effort under arbitrary (non-parametric) finite type-spaces remains open. It would also be interesting to study variants of the design problem that allow for the possibility of more general mechanisms, a direction recently explored by \citet*{letina2023optimal} and \citet*{zhang2024optimal} in the complete information and continuum type-space environments, respectively. Furthermore, we believe the finite type-space environment provides a more convenient framework for experimental investigations compared to the continuum type-space. With the equilibrium predictions and convergence properties we establish, we hope to also inspire more experimental research investigating some of the theoretical predictions in the literature on contest design with incomplete information.




\newpage
\bibliographystyle{ecta}

\bibliography{refs}

\end{document}